\documentclass[aps,prl,reprint]{revtex4-2}
\usepackage{hyperref}
\usepackage{amsmath}
\usepackage{amssymb}
\usepackage[pdftex]{graphicx}
\usepackage{microtype}
\usepackage{verbatim}
\usepackage{todonotes}
\usepackage{subcaption}
\usepackage{multirow}
\usepackage{array,booktabs,tabularx,tabulary}
\usepackage{adjustbox}
\usepackage{placeins}
\makeatletter
\DeclareRobustCommand{\element}[1]{\@element#1\@nil}
\def\@element#1#2\@nil{%
  #1%
  \if\relax#2\relax\else\MakeLowercase{#2}\fi}
\makeatother

\newcommand{\mrm}{\mathrm}

\newcommand{\ket}[1]{{\left| {#1} \right\rangle}}
\newcommand{\bra}[1]{{\left\langle {#1} \right|}}

\begin{document}
\title{BaF molecules in neon ice: trapping, spectroscopy and optical control of electron spins}
\author{S. J. Li}
\author{H. D. Ramachandran}
\author{R. Anderson}
\author{A. C. Vutha}
\affiliation{Department of Physics, University of Toronto, Toronto ON M5S 1A7, Canada}

\begin{abstract}
We have trapped BaF molecules in neon ice, and used laser-induced fluorescence spectroscopy to map out optical transitions in the trapped molecules. Our measurements show that the neon lattice does not significantly perturb certain optical transitions in the trapped molecules. We used one of these transitions to polarize the electron spins, detect spin flips and measure hyperfine transitions in the trapped molecules, entirely using lasers. This demonstration with heavy polar molecules opens up new opportunities for precision measurements of beyond-standard-model physics.
\end{abstract}

\maketitle
The inability of the standard model of particle physics to explain cosmological observations, such as the abundance of dark matter and the matter-antimatter asymmetry of the universe, motivates the search for new physics that lies beyond the reach of present-day colliders. Measurements using atoms and molecules offer a means to probe such beyond-standard-model (BSM) physics, due to the intrinsically high sensitivity of certain heavy atoms and molecules, and the extreme precision with which anomalous frequency shifts can be measured using atomic physics techniques \cite{DeMille2017,Safronova2018}. Therefore precision searches for new physics commonly require the ability to prepare chosen atoms and molecules in specific quantum states, and to detect transitions from these states with high signal-to-noise ratio (SNR). 

The need for high precision naturally drives the use of large ensembles of atoms or molecules. Further, trapping the atoms or molecules allows for repeated measurements on the ensemble and control of motion-derived systematic errors. Large trapped ensembles offering high SNR are useful for precision measurements of static BSM signals (e.g., permanent electric dipole moments, EDMs \cite{Singh2019,Vutha2018}), but they can also be valuable in searches for transient or time-dependent effects (e.g., domain wall dark matter \cite{Derevianko2014} or oscillating EDMs \cite{Graham2011,Stadnik2014}) where high precision must be attained within a short duration. In this context the technique of matrix isolation -- where atoms and molecules are frozen within cryogenically-grown inert ices -- offers a simple way to trap a wide variety of particles \cite{Weltner1978,Kanorsky1996,Bondybey1996,Dargyte2021,Lancaster2021,Lambo2021}. This technique is particularly attractive as a way to trap heavy molecules \cite{Knight1971,Weltner1995} without the challenges associated with laser-cooling \cite{Zhang2022}. Large samples can be obtained, containing orders-of-magnitude more trapped molecules than possible with any other method. Crucially, the interaction of the inert matrix with the molecules can be weak enough to avoid distorting their properties, while at the same time providing sufficiently strong confinement that molecules can be cooled to their quantum ground state of motion using straightforward cryogenic means.

The physical chemistry of matrix-isolated molecules has been studied using magnetic resonance (MR) techniques for many years \cite{Weltner1995,Bondybey1996}, demonstrating that electron and nuclear spins in these molecules can be coherently manipulated. However, significant obstacles remain to be confronted before the full potential of matrix-isolated molecules can be exploited for precision measurements of BSM physics. First, many precision measurements (e.g., EDM searches \cite{Andreev2018}) operate at low magnetic fields: thus traditional MR methods for creating spin polarization, using tesla-scale magnetic fields, are infeasible, leaving optical pumping as one of the few viable options. Second, it is desirable to detect state transitions in trapped molecules with high SNR -- higher than possible with the thermal-noise-limited detectors used in traditional MR -- in order to take full advantage of the BSM sensitivity of large ensembles of molecules. Again, optical techniques such as laser-induced fluorescence enable high-SNR detection. But before one can begin to address either of these challenges using lasers, there remains a third obstacle, which is that there are no published high-resolution optical spectra in inert ices for any of the molecules that are of interest for BSM searches. 

In this work we systematically clear these obstacles: after showing that high-resolution laser spectra can be obtained from molecules trapped in inert ice, we use one of the newly-identified optical transitions to spin-polarize electrons in the molecules, read out spin state transitions, and perform precision hyperfine spectroscopy using a laser. We demonstrate these advances using barium monofluoride molecules trapped in neon ice (BaF:Ne). Besides considerations of experimental convenience (e.g., near-infrared optical transitions), we focused on BaF due to its role in current \cite{Aggarwal2018} and proposed \cite{Vutha2018} searches for new physics. Based on the physics uncovered during this investigation, we surmise that the methods and results described here can be generalized to other molecules of interest.

\begin{figure}[h] 
\includegraphics[width=\columnwidth]{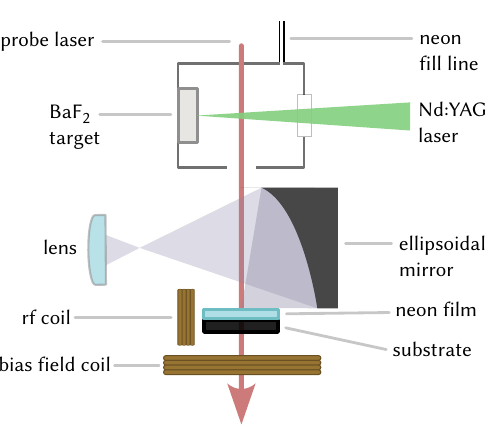}
\caption{Schematic of the apparatus used for the measurements. A neon film grown on a cryogenic sapphire substrate was doped with BaF molecules produced by laser ablation in a buffer-gas cell. Fluorescence from the molecules in neon was collected using an ellipsoidal mirror and lens, and coupled to a detector outside the cryostat.}
\label{fig:apparatus}
\end{figure}

\emph{Methods. -- } BaF:Ne samples were grown by co-depositing BaF molecules and neon atoms onto the surface of a cooled sapphire disk. A schematic diagram of the apparatus is shown in Fig.\ \ref{fig:apparatus}. 
We used a pulsed Nd:YAG laser to ablate a ceramic BaF$_2$ disk, held within a neon buffer gas cell maintained at 20 K. A steady flow of 20 standard cm$^3$/minute of neon gas was sent through the cell during the BaF:Ne sample growth process. BaF molecules produced from ablation were cooled in collisions with the neon atoms, carried by the gas flow exiting the cell and deposited 30 mm downstream onto the sapphire substrate along with the neon atoms. Absorption of a laser tuned to the gas-phase BaF $X \, {}^2\Sigma, v=0, J=1/2^+ \to A \, {}^2\Pi_{1/2}, v=0, J=3/2^-$ 
transition was used to monitor the density of molecules in the buffer gas cell. BaF:Ne samples were typically grown at a rate of 18 $\mu$m/min for 10 minutes, after which the laser ablation and gas flow were stopped and the temperature of the substrate lowered to 5 K for further measurements. A SmCo magnet and a set of coils were used to apply a static magnetic field (0-100 G) normal to the substrate. A second coil was used to generate an rf magnetic field in the plane of the substrate for spin resonance experiments. 

Laser-induced fluorescence measurements were made using a tunable Ti:sapphire laser and external-cavity diode lasers. An electro-optic modulator was used to control the amplitude of the lasers and to switch it on or off rapidly (in $<$ 10 ns). A waveplate on a motorized stage was used to control the polarization of the lasers. We typically used 3 mW of laser light focused onto a 30-$\mu$m-diameter spot on the neon film, although experiments were performed at a range of powers between 0.01-100 mW to ensure that the results reported here were not affected by bleaching or saturation. Laser-induced fluorescence from the illuminated region was collected using an ellipsoidal mirror and a lens, coupled out of the cryogenic vacuum system through a viewport, and sent through bandpass filters into a detector (either a spectrometer, camera or avalanche photodiode depending on the measurement).

\begin{figure*}[ht!]
\includegraphics[width=\textwidth]{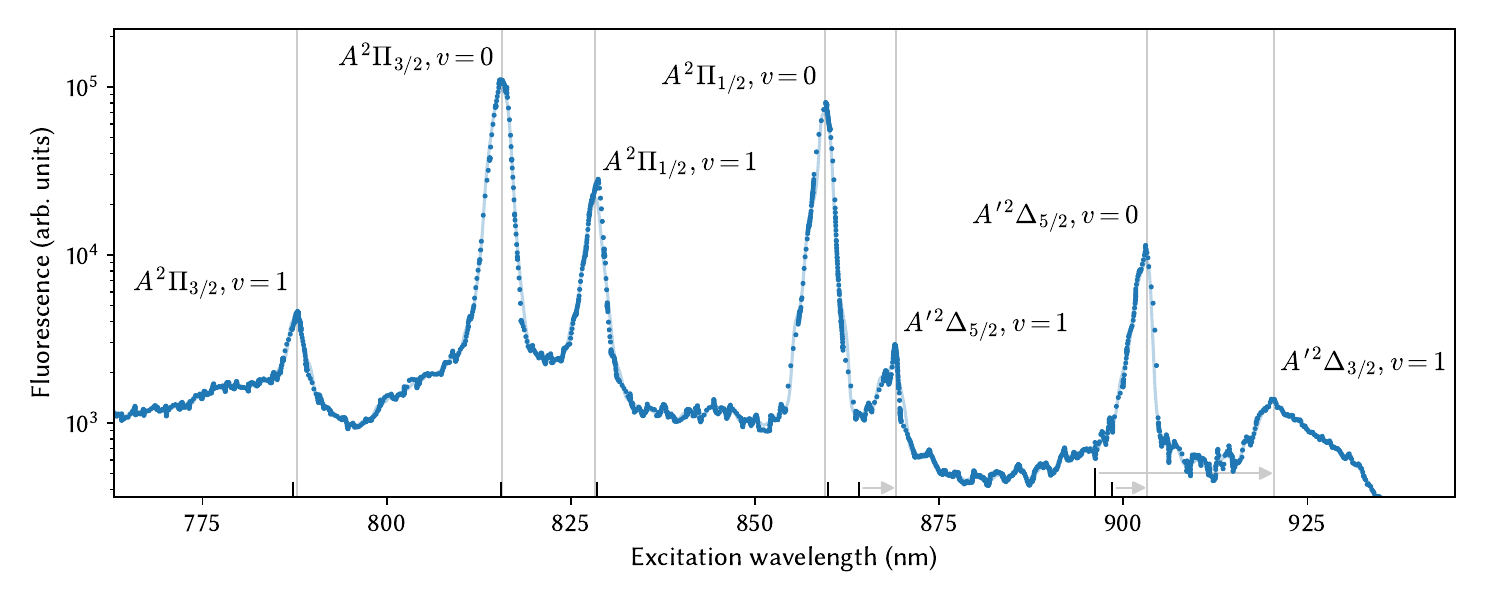}
\caption{Laser-induced fluorescence from BaF:Ne as a function of the wavelength of the excitation laser. The signal on the vertical axis is the integrated fluorescence emission between 950-970 nm, normalized by the excitation laser power at each wavelength. Note the logarithmic scale. Labels accompanying the vertical lines through the peaks denote states excited out of the ground $X \, {}^2\Sigma, v=0$ state, whose corresponding gas-phase positions are indicated by bars on the horizontal axis.}
\label{fig:excitation_spectrum}
\end{figure*}    
\emph{Results. -- } Dispersed fluorescence from BaF:Ne was measured on a spectrometer as the probe laser was tuned over approximately 150 nm. Emission between 950-1100 nm was observed following excitation of the BaF:Ne at a number of distinct wavelengths (see Supplemental Material, SM). Fluorescence from a single BaF:Ne sample remained stable and repeatable over many weeks of measurements. 

Fluorescence emission in the 950-970 nm band is shown plotted as a function of the wavelength of the excitation laser in Fig.\ \ref{fig:excitation_spectrum}. The spectrum reveals a distinct fingerprint of BaF molecules in neon. The peaks in Fig.\ \ref{fig:excitation_spectrum} were identified based on their close coincidence with spectral lines in gas-phase BaF molecules \cite{Steimle2011,Chen2016}. The observed line centers and linewidths are listed in Table I in the SM. Using the measured geometric and quantum efficiencies of our detection system, we estimate a fluorescence emission rate of $F_\mrm{emission} \approx 10^{13}$ photons/s/mW at the peak of the $X,v=0 \to A \, ^2\Pi_{1/2},v=0$ resonance. 

The spatial electron distributions, bond lengths and dipole moments of the $A' \, {}^2\Delta$ states in BaF are rather different from those of the $X \, {}^2\Sigma$ or $A \, {}^2\Pi$ states \cite{Hao2019}. It is possible that these properties are responsible for stronger coupling of the $A'$ states to the neon atoms around a trapped BaF molecule, consistent with our observation that deviations from gas-phase lines were much larger for the $A' \, ^2\Delta_{5/2}$ ($\approx 58$ cm$^{-1}$) and $A' \, ^2\Delta_{3/2}$ states ($\approx 300$ cm$^{-1}$) states compared to the $A \, {}^2\Pi$ states ($< 7$ cm$^{-1}$). Non-radiative quenching may also be responsible for the fact that fluorescence is predominantly emitted in the $A'\, {}^2\Delta_{3/2} \to X \, {}^2\Sigma$ band after excitation of the $X \, {}^2\Sigma \to A \, {}^2\Pi$ transitions. Larger shifts and broadening compared to the gas phase have been reported for transitions to higher-lying excited states of BaF in neon \cite{Knight1971}, consistent with the above picture. 

The spectrum of BaF molecules in neon presents an interesting contrast to alkali atoms in neon \cite{Dargyte2021,Lancaster2021}, where the $s \to p$ transitions were found to be broadened and shifted (by over 100 nm in some cases) compared to the gas phase. In BaF molecules, which are isoelectronic to alkali atoms, the analogous $X \to A$ transitions are significantly narrower and nearly unshifted from their gas-phase counterparts. This fact points to a very different local environment for molecules in neon ice compared to atoms. The qualitative difference between the spectra of trapped atoms and molecules has in the past been attributed to the diminished influence of the neon lattice on the valence electrons in molecules compared to atoms, since the valence electrons are partially involved in the molecular bond \cite{Bondybey1978}. We posit that such an explanation cannot be the entire story, since the valence electron in BaF occupies a non-bonding orbital that is decoupled from the Ba-F bond. Therefore, it is remarkable that the molecular transitions shown in Fig. \ref{fig:excitation_spectrum} are resolvable and relatively narrow. The small spectral shifts of the $X \to A$ transitions imply that the electronic wavefunctions of the $X$ and $A$ states are not strongly perturbed by the neon lattice. The vibrational constants in the $A \, {}^2\Pi$ states as observed from the vibrational satellite lines in Fig.\ \ref{fig:excitation_spectrum} are also consistent with their gas-phase values, suggesting that the Ba-F stretch mode is not significantly perturbed by the neighboring Ne atoms either.

\begin{figure}[h!]
\includegraphics[width=\columnwidth]{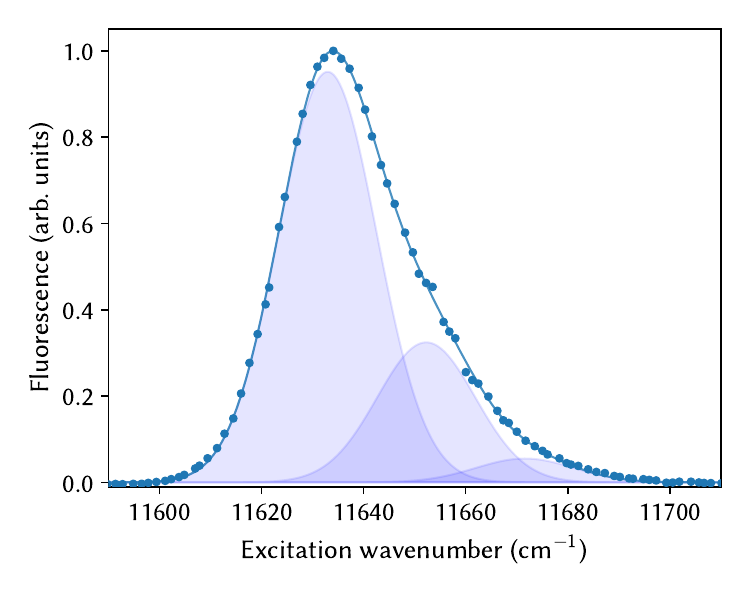}
\caption{High-resolution spectrum of the $X \, {}^2\Sigma, v=0 \to A \, {}^2\Pi_{1/2}, v=0$ transition. The solid line through the data points is a fit to a zero-phonon line plus phonon sidebands (component lines shown shaded) using the model described in the text.
}
\label{fig:860_fine}
\end{figure}     

High-resolution spectroscopy of the BaF:Ne lines provides insights into the environment of the trapped molecules. For instance, the high-resolution spectrum of the $X,v=0 \to A \, ^2\Pi_{1/2},v=0$ line in Fig.\ \ref{fig:860_fine} shows a distinctly asymmetric lineshape. We model this effect as a consequence of coupling between a molecule and a local phonon mode, caused by the change in the valence electron distribution during the optical transition. We fit the observed lineshape to the function $f(x) = \sum_{m=0}^2 \delta(x - m \nu_p) \, e^{-S} \, S^m/m! $ convolved with a gaussian with variance $\sigma^2$ centered at $\nu_0$, the frequency of the zero-phonon line. Here $S$ is the Huang-Rhys factor that parameterizes the strength of the coupling between the molecules and the mode, $\nu_p$ is the frequency of the mode and $m$ is the number of quanta in the mode. The data agree well with this simple model, as shown in Fig.~\ref{fig:860_fine}, and the fit to the model gives $\nu_0$ = 11632.96(9) cm$^{-1}$, $\nu_p$ = 19.3(2) cm$^{-1}$ and $\sigma$ = 9.6 cm$^{-1}$. The Huang-Rhys factor, $S = \bar{m} = 0.34(1)$, implies weak coupling between the molecule and the mode, and an appreciably high probability that the molecule remains in its motional quantum ground state while the valence electron is optically excited. \textit{Ab initio} calculations combined with this measurement of the coupling strength and $\nu_0$ could identify, e.g., whether this mode is due to the libration of oriented BaF molecules or their center-of-mass motion within a neon cage.
       
\begin{figure}
\includegraphics[width=\columnwidth]{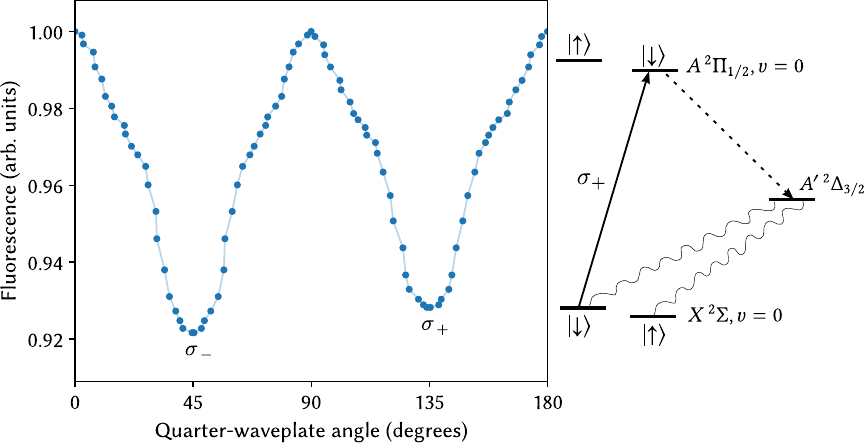}
\caption{Fluorescence following laser excitation of the $X \, {}^2\Sigma \to A \, {}^2\Pi_{1/2}$ transition, as the laser polarization was tuned using a quarter-waveplate. The line through the data points is a guide to the eye. $\sigma_+ (\sigma_-)$ denotes right (left) circular polarization. The dips in fluorescence for circularly polarized light are due to optical pumping of the molecules into spin states that are dark to the laser, as shown in the schematic level diagram.} 
\label{fig:optical_pumping}
\end{figure}   
The optical transitions observed in this work make it possible to control the electron spins in BaF:Ne using entirely optical means. We used the $X \, {}^2\Sigma,v=0 \to A \, {}^2\Pi_{1/2},v=0$ transition (hereafter referred to as ``$D1$'') to optically polarize the BaF valence electron spins, and then optically detect their precession in a magnetic field. Figure \ref{fig:optical_pumping} shows the dependence of fluorescence from a laser with intensity $I = 2.4$ kW/cm$^2$ on the laser polarization. A static magnetic field $\vec{B} = B_z \hat{z}$ with $B_z$ = 20 G was applied during this measurement. The polarization of the laser was varied between linear (0$^\circ$, 90$^\circ$, 180$^\circ$) and left (45$^\circ$, $\sigma_-$) or right (135$^\circ$, $\sigma_+$) circular by adjusting the angle of a quarter-waveplate. The dependence of fluorescence on the laser polarization was corrected for the birefringence of in-vacuum optics (see SM, Sec.\ C). The decrease in fluorescence observed for circularly versus linearly polarized light indicates optical pumping of the molecules into spin states that do not fluoresce in circularly polarized light.

\begin{figure}[h!]
\includegraphics[width=\columnwidth]{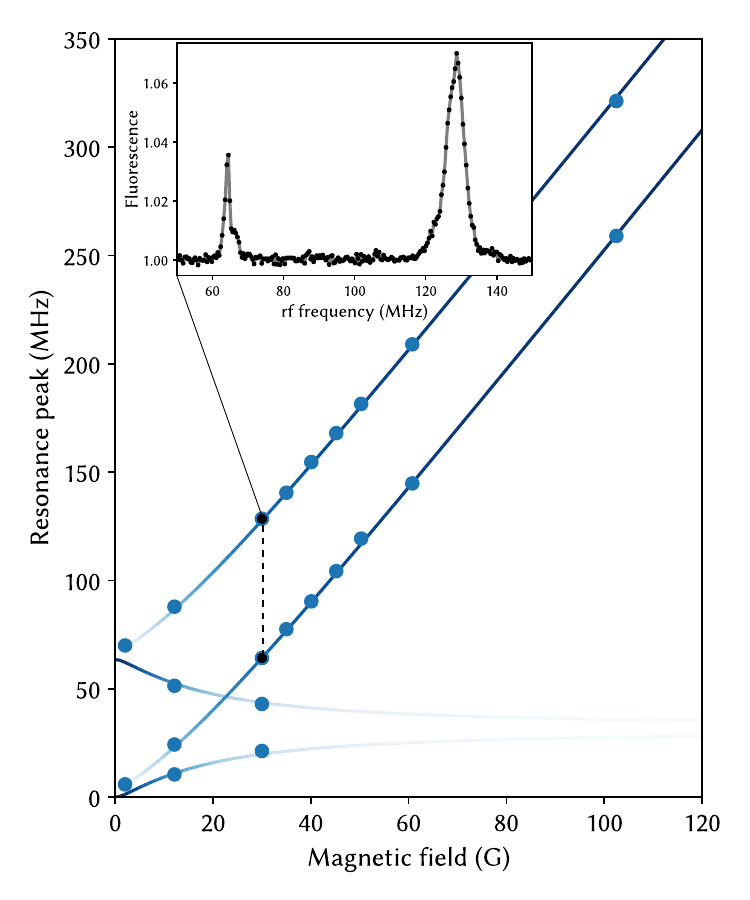}
\caption{Resonances in the hyperfine spectrum for various values of the static $B$-field. Lines through the data points are the calculated hyperfine transition frequencies for BaF molecules oriented parallel to the substrate, and their boldness represents the calculated transition strength (see SM for details of the calculations). The inset shows a section of the optically-detected hyperfine spectrum at $B_z$ = 30 G.}
\label{fig:esr}
\end{figure}     
To confirm the creation of spin polarization, we illuminated the BaF:Ne sample with $\sigma_+$ $D1$ light, applied a continuous-wave rf magnetic field $\vec{B}_\mrm{rf} = B_x(t) \hat{x}$ perpendicular to the static field, and varied the rf frequency. When the rf $B$-field resonantly drives hyperfine transitions in BaF:Ne, repopulation of the dark states leads to revival of laser-induced fluorescence. The result therefore is an optically-detected spin resonance spectrum, as shown in the inset to Fig.~\ref{fig:esr}. 

The positions of peaks in the optically-detected spin resonance spectrum are plotted against static $B$-field values in Fig.\ \ref{fig:esr}. The solid lines in the figure are calculated transition frequencies for molecules whose internuclear axis is fixed and perpendicular to the static field (see SM, Sec.\ D). The clear agreement of the observed resonances with this calculation is strong evidence that the trapped BaF molecules are non-rotating and aligned parallel to the substrate. We note that similar orientation-locking effects have been reported in microwave electron spin resonance measurements on matrix-isolated molecules (e.g., \cite{Knight1971a,Brom1972,Weltner1995}). 

\begin{figure}[h!]
\includegraphics[width=\columnwidth]{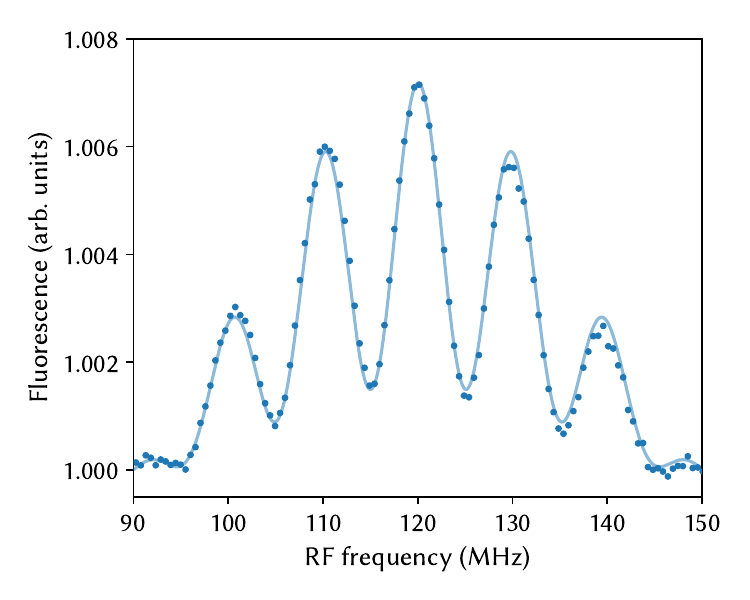}
\caption{Optically-detected hyperfine resonance. For this measurement the $\sigma_+$-polarized $D1$ laser was pulsed on and off at 20 kHz, and two 30-ns-long rf pulses with a separation of 100 ns were applied during the dark phase. The solid line is a fit to a Ramsey lineshape, which yields an uncertainty of 10 kHz in the center frequency.}
\label{fig:ramsey}
\end{figure}     

Equipped with means to optically polarize the BaF electron spins and read them out with high SNR, we performed pulsed rf spectroscopy of BaF:Ne. Figure \ref{fig:ramsey} shows an optically-detected hyperfine resonance using the double-pulse Ramsey method, measured to 10 kHz precision in 300 s. This measurement establishes that molecules trapped in inert ices can be spin-polarized in low magnetic fields and their spin transitions measured using lasers. 

In summary, we have demonstrated stable trapping of BaF molecules in neon ice, mapped out their laser-induced fluorescence spectrum, and performed optically-controlled spin resonance measurements. In the process we have discovered nearly-unperturbed optical transitions and evidence that trapped molecules are locked in fixed orientations within neon ice. We anticipate that our work will unlock further investigations into the physics of trapped molecules in inert ices, and spark novel applications of trapped molecules to searches for new physics.

~\\~\\
\emph{Acknowledgments. --} We have enjoyed helpful conversations with Jonathan Weinstein, Alexei Buchachenko, Tim Steimle and Andrew Jayich. Mohit Verma, Camilo Sanchez, Jessica Patel and Eman Shayeb contributed to the development of the apparatus. We acknowledge discussions with Eric Hessels, Jaideep Singh and other members of the EDM$^3$ Collaboration, which is funded by the Natural Sciences and Engineering Research Council of Canada, Canada Foundation for Innovation, Ontario Research Fund, Northwestern Center for Fundamental Physics, Gordon and Betty Moore Foundation, and Alfred P. Sloan Foundation. SJL acknowledges support from an NSERC undergraduate research award. ACV acknowledges support from the Canada Research Chairs program.
    
\bibliography{bafne.bib}

\appendix    
\onecolumngrid 
\newpage
\section*{Supplemental Material}

\subsection*{A. Line centers and widths}
\begin{table}[h!]
    \centering
    \begin{tabular}{lllll}
    \toprule
    transition & wavelength (nm) & wavenumber (cm$^{-1}$) & FWHM (cm$^{-1}$) & shift  (cm$^{-1}$) \\
    \midrule
    $X,v=0 \to A\,^2\Pi_{3/2},v=1$ & 787.66(5) & 12692.3(8) & 25.2(19) & -6.2(8) \\
    $X,v=0 \to A\,^2\Pi_{3/2},v=0$ & 815.572(2) & 12257.97(3) & 28.60(8) & -3.88(3) \\
    $X,v=0 \to A\,^2\Pi_{1/2},v=1$ & 828.302(6) & 12069.57(9) & 24.3(2) & 4.12(9) \\
    $X,v=0 \to A\,^2\Pi_{1/2},v=0$ & 859.385(3) & 11633.03(5) & 23.15(9) & 3.08(5) \\
    $X,v=0 \to A'\,^2\Delta_{5/2},v=1$ & 868.52(6) & 11510.6(8) & 18.3(15) & -58.0(8) \\
    $X,v=0 \to A'\,^2\Delta_{5/2},v=0$ & 902.73(4) & 11074.4(4) & 22.8(9) & -55.3(4) \\
    $X,v=0 \to A'\,^2\Delta_{3/2},v=1$ & 920.9(7) & 10856(8) & 84(23) & -298(8) \\
    \bottomrule
    \end{tabular}
    \caption{Center wavelengths (in air), center wavenumbers (in vacuum), and full-widths at half maximum for the transitions in BaF:Ne shown in Fig. \ref{fig:excitation_spectrum}. The last column lists the difference between the observed lines and the corresponding gas-phase BaF transitions. }
    \label{tab:lines}
\end{table}

\subsection{B. Fluorescence spectrum and dynamics}
\begin{figure}[h!]
\includegraphics[width=0.6\columnwidth]{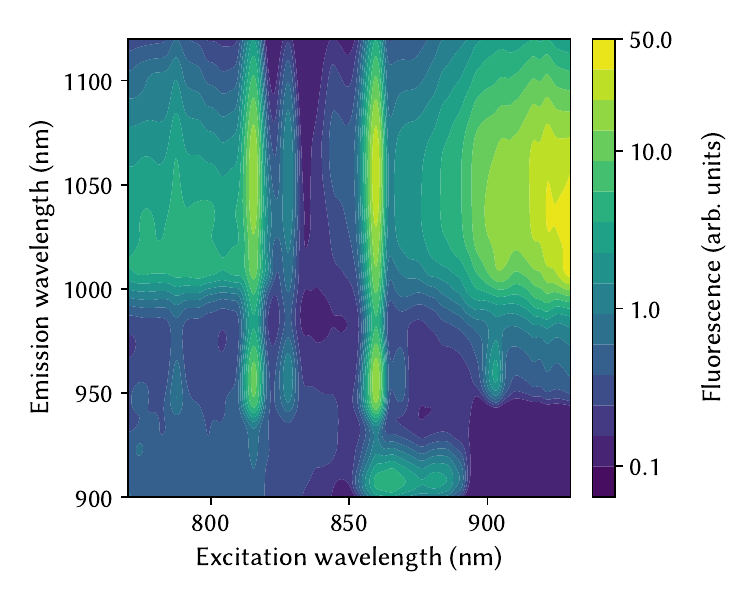}
\caption{Dispersed fluorescence from BaF molecules in neon. The color bar shows the intensity of the fluorescence on a logarithmic scale. The vertical features between approximately 950-1100 nm correspond to phonon-broadened emission from the $A' \, {}^2\Delta_{3/2}$ state following excitation of various BaF lines.}
\label{fig:dispersed_fluorescence}
\end{figure}    

Fig.\ \ref{fig:dispersed_fluorescence} shows the dispersed fluorescence obtained for a range of excitation laser wavelengths. A horizontal section through Fig.\ \ref{fig:dispersed_fluorescence}, integrated between 950-970 nm, leads to the excitation spectrum plotted in Fig.\ \ref{fig:excitation_spectrum}. 

\begin{figure}[h!]
\includegraphics[width=0.6\columnwidth]{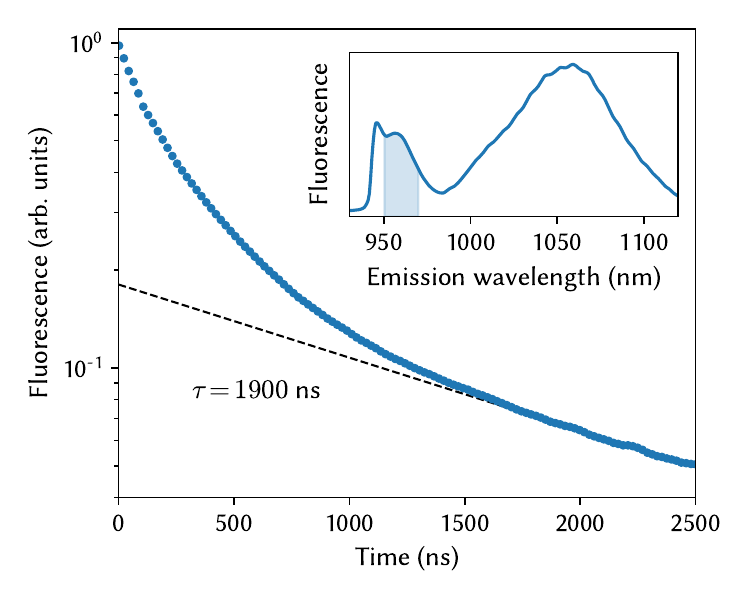}
\caption{Time dependence of fluorescence emission, following excitation of the $X \, {}^2\Sigma, v=0 \to A \, {}^2\Pi_{1/2}, v=0$ transition at 859.4 nm. The inset shows the emission spectrum, and the shaded 950-970 nm band indicates the spectral window over which the decay time dependence was measured. The zero of time denotes the instant when the excitation laser was switched off. The decay is not consistent with a single exponential, suggesting that phonon-assisted mechanisms may be at work.
The eventual exponential decay at long times (dashed line) has a $1/e$ time $\tau=1900$ ns.}
\label{fig:delta_lifetime}
\end{figure}

We attribute the 950-1100 nm emission features in Fig.\ \ref{fig:dispersed_fluorescence} to decays from the $A' \, {}^2\Delta_{3/2}$ state to the ground electronic $X \, {}^2\Sigma$ state, broadened by phonons due to coupling between the $A'$ state and the neon lattice. We measured the dynamics of emission from the $A' \, {}^2\Delta_{3/2}$ state by monitoring fluorescence after the excitation laser was switched off. The resulting fluorescence decay curve is shown in Fig.~\ref{fig:delta_lifetime}. 
\begin{figure}[h!]
 \includegraphics[width=0.6\columnwidth]{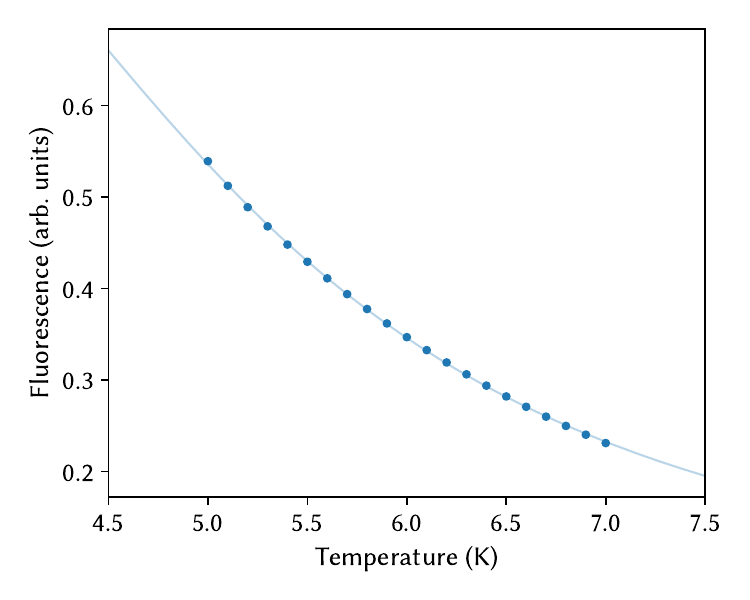}
 \caption{Laser-induced fluorescence in the 950-970 nm band after resonant excitation of the $X \, ^2\Sigma,v=0 \to A \, {}^2\Pi_{1/2},v=0$ transition, as a function of the substrate temperature. 10 mW of laser power was used during this measurement -- there was no measurable temperature change due to the laser. The solid line is a fit to the nonradiative decay model in Eq.~\ref{eq:quenching}.}
 \label{fig:thermal_quenching}
\end{figure}     
The decay is distinctly non-exponential up to 1500 ns, indicating that a simple radiative decay from the $A' \, {}^2\Delta_{3/2}$ state is not the only mechanism involved. The observed dependence of the fluorescence rate on temperature in Fig.\ \ref{fig:thermal_quenching} is also suggestive of a thermally-activated nonradiative decay pathway. We assume that a nonradiatively decaying state with lifetime $\tau_{nr}$ and activation energy $E_a$ competes with a radiative decay pathway that has lifetime $\tau_r$, resulting in an overall decay rate $\tau^{-1} = \tau_r^{-1} + e^{-E_a/k_B T} \tau_{nr}^{-1}$. Then the branching ratio for radiative decay, $\epsilon_r$, varies as
\begin{equation}
\epsilon_r = \left(1 + \frac{\tau_r}{\tau_{nr}} e^{-E_a / k_B T}\right)^{-1}.
\label{eq:quenching}
\end{equation}
This model fits well to the observed temperature dependence shown in Fig.~\ref{fig:thermal_quenching}, with $E_a = hc \times 16.2(3)$ cm$^{-1}$ and $\tau_r/\tau_{nr} = 94(2)$.

\subsection*{C. Birefringence correction}
\begin{figure}[h!]
\includegraphics[width=0.5\textwidth]{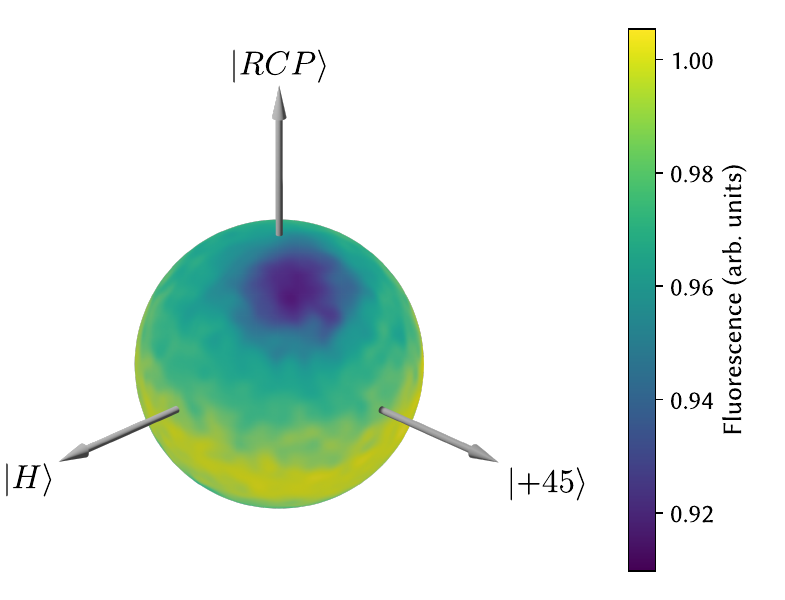}
\caption{Fluorescence rate as a function of polarization state, visualized over the surface of the Poincar\'e sphere. The two antipodal minima in fluorescence are tilted with respect to the vertical due to residual birefringence of in-vacuum optics.}
\label{fig:poincare_sphere}
\end{figure}

To obtain the data shown in Fig.~\ref{fig:optical_pumping},
we used two waveplates to sample uniformly over all possible polarization states of the excitation beam.
Fig.~\ref{fig:poincare_sphere}
shows the fluorescence rate as a function of polarization state, visualized over the surface of the Poincar\'e sphere.
There are two antipodal minima in fluorescence rate, corresponding to pure circular polarization at the location of BaF:Ne sample. A small amount of birefringence in the in-vacuum optics tilts these points away from the vertical axis.

After correcting for this tilt, we average the fluorescence rate azimuthally over the Poincar\'e sphere. This process eliminates the effect of the angular distribution of emitted fluorescence for the following reason. The emission pattern from a classical dielectric scatterer follows the angular distribution
\[
I_{CP}(\theta, \phi) \propto \frac{1}{2} (1 + \cos^2{\theta})
\]
when illuminated with circularly polarized light propagating along the $z$-axis,
and
\[
I_{x}(\theta, \phi) \propto \sin^2{\theta} \sin^2{\phi} + \cos^2{\theta}
\]
when illuminated with $x$-polarized light.
Here $\theta, \phi$ are the polar and azimuthal angles, respectively, in a $z$-aligned spherical coordinate system.
Thus the efficiency of the fluorescence collection optics will vary depending on the incident light polarization.
However, observe that
\[
\int_0^{2\pi} I_{CP}(\theta, \phi) \,\mathrm{d}{\phi}
= \int_0^{2\pi} I_{x}(\theta, \phi) \,\mathrm{d}{\phi},
\]
so performing an azimuthal average over the Poincar\'e sphere eliminates the polarization dependence. After this procedure, any variations in the averaged signal are due to changes in the \textit{total} emission rate as a function of polarization.

\subsection{D. Hyperfine resonances of oriented molecules}
The hyperfine structure of a BaF molecule in the ground electronic $X \, {}^2\Sigma$ state is described by the effective Hamiltonian
\[
H = b \vec{I} \cdot \vec{S} + c I_k S_k - g \mu_B B_z S_z,
\]
where $b = h \times 63.5(32)$ MHz and $c = h \times 8.224(58)$ MHz are hyperfine constants for BaF \cite{Ernst1986} and $\vec{B} = B_z \hat{z}$ is the static magnetic field. Here $\hat{k}$ is the direction of the molecular axis. The spectrum of this hyperfine Hamiltonian depends on the angle $\theta$ between $\hat{k}$ and $\hat{z}$. The transition frequencies calculated by diagonalizing the Hamiltonian with $\theta = 90^\circ$ (molecular axis parallel to the surface of the sapphire substrate) are in excellent agreement with the experimentally measured resonances, as shown in Fig.\ \ref{fig:esr}. 

The effective Hamiltonian can also be used to understand the strengths of the transitions. In the high-field limit $\mu_B B_z \gg b, c$, the eigenstates of $H$ are uncoupled basis states $\ket{m_S, m_I} = \ket{\downarrow \downarrow}, \ket{\downarrow \uparrow}, \ket{\uparrow\downarrow}, \ket{\uparrow \uparrow}$. Let $\ket{0}, \ket{1}, \ket{2}, \ket{3}$ denote the eigenstates that are adiabatically connected to these high-field states, in this order. In our experiment the rf field coil generates an oscillating transverse magnetic field $\vec{B}_\mrm{rf} = B_x(t) \hat{x}$, which couples to the spin component $S_x$. When the angle $\theta = 90^\circ$, the matrix elements $\bra{3} S_x \ket{0}$ and $\bra{2} S_x \ket{1}$ vanish identically. Moreover, the dipole couplings $\bra{1} S_x \ket{0}$ and $\bra{3} S_x \ket{2}$ decrease as $\sim b / (4 g \mu_B B_z)$ for $\mu_B B_z \gg b$. The remaining two transitions $\ket{0} \to \ket{2}$ and $\ket{1} \to \ket{3}$ have order-unity matrix elements of $S_x$ for all values of $B_z$. Therefore for molecules with $\theta = 90^\circ$ we expect to observe four transitions, two of which become undetectable in the high-field limit. Thus the calculated strength of the observed transitions is also consistent with the behavior observed in Fig.\ \ref{fig:esr}.
 
\end{document}